\documentclass[aps,prl,a4paper,showpacs,twocolumn,floatfix,groupedaddress,10pt]{revtex4}

\usepackage{amsfonts}
\usepackage{amssymb}
\usepackage{amsmath}
\usepackage{graphicx}
\usepackage{dcolumn}
\usepackage{bm}

\usepackage{enumerate}
\usepackage{amsmath}

\newcommand{\ket}[1]{\left|#1\right\rangle}

\newcommand{\ketbra}[1]{\ket{#1}\left\langle #1\right|}

\begin{document}


\title{Towards Communication-Efficient Quantum Oblivious Key Distribution}

\author{M. V. Panduranga Rao$^1$, M. Jakobi$^2$}
\affiliation{%
$^1$Department of Computer Science and Engineering
Indian Institute of Technology Hyderabad\\
$^2$Group of Applied Physics, University of Geneva, CH-1211 Geneva 4, Switzerland
}%

\date{\today}

\begin{abstract}
Oblivious Transfer, a fundamental problem in the field of secure multi-party computation is defined as follows: A database $D\!B$ of $N$ bits held by Bob is queried by a user Alice who is interested in the bit $D\!B_b$ in such a way that (1) Alice learns $D\!B_b$ and only $D\!B_b$ and (2) Bob does not learn anything about Alice's choice $b$. While solutions to this problem in the classical domain rely largely on unproven computational complexity theoretic assumptions,
it is also known that perfect solutions that guarantee both database and user privacy are impossible in the quantum domain.

Jakobi et al. [Phys. Rev. A, 83(2), 022301, Feb 2011] proposed a protocol for Oblivious Transfer using well known QKD techniques to establish an \emph{Oblivious Key} to solve this problem. Their solution provided a good degree of database and user privacy (using physical principles like impossibility of perfectly distinguishing non-orthogonal quantum states and the impossibility of superluminal communication) while being loss-resistant and implementable with commercial QKD devices (due to the use of SARG04).

However, their Quantum Oblivious Key Distribution (QOKD) protocol requires a communication complexity of $O(N \log N)$. Since modern databases can be extremely large, it is important to reduce this communication as much as possible.

In this paper, we first suggest a modification of their protocol wherein the number of qubits that need to be exchanged 
is reduced to $O(N)$. A subsequent generalization reduces the quantum communication complexity even further in such a way that only a few hundred qubits are needed to be transferred even for very large databases.

\end{abstract}
\maketitle

\section{Introduction\label{sec:level1}}

Impressive progress has been made over the last two decades in our understanding of how Quantum principles can be used to secure communication between trustful parties against eavesdropping. For example, Quantum Key Distribution (QKD) techniques have gained steadily in technical applicability. However, in the more general field of secure multi-party computation, which comprises tasks such as Coin Flipping and Bit Commitment and normally implies communication between distrustful parties, only a few quantum alternatives to classical schemes have emerged.
One of the most fundamental problems of this type is Oblivious Transfer (OT), also known as Symmetrically Private Information Retrieval (SPIR). This task is complete for secure multi-party computations in the sense that all other tasks may be constructed from it~\cite{Kilian}.
Originally introduced in two different flavors by Rabin~\cite{Rabin} in 1981 and by Even, Goldreich and Lempel~\cite{EGL} in 1985, which were shown to be equivalent by Cr\'{e}peau~\cite{Crepeau}, the problem of 1-out-of-2 OT requires Bob to send two bits to Alice such that (i) Alice gets to receive only one bit -- she cannot get significant
information about the other -- and (ii) Bob does not get to know which bit Alice received, i.e. he is oblivious to what she learns.
The problem of 1-out-of-N OT is a generalization of the 1-out-of-2 OT: Bob hosts a database $D\!B$ of $N$ bits. Alice wishes to retrieve the value of a certain bit, say the $b^{th}$, from the database. Privacy
has to be preserved symmetrically: Bob should not get to know which bit Alice is interested in (that is, in 
this case he should not get to know $b$); at the same time, Alice should not get to know the value of
any other bit in the database that she has not queried for.

It is interesting to note that this task may accomplished by precomputing an ``Oblivious Key''~\cite{OK}: a string $O\!K$ of $N$ random bits that is (i) completely known to Bob while (ii) Alice knows only one bit $O\!K_j$ of this string, with Bob being oblivious to $j$. Once such a key is established, it can be used to complete the actual OT: Alice being interested in the database element $D\!B_b$ announces a shift $s=j-b$ to Bob. Thus, Bob gets to know neither $j$ nor $b$, but only $s$. 
Bob then encrypts the database bitwise as $D\!B'_a = \mathrm{X\!O\!R}(D\!B_a,O\!K_{a+s})$, $1\leq a\leq N$, and announces the encrypted database $D\!B'$. From this, Alice recovers the bit that she wanted: $D\!B_b=\mathrm{X\!O\!R}(D\!B'_b,O\!K_j)$ and the OT is complete.

There exist several approaches in the classical realm to SPIR and OT (see e.g.~\cite{Kilian,jacmChor,Rabin,EGL}).
Existing classical protocols for these problems depend on some unproven computational
complexity theoretic assumption like nonexistence of efficient algorithms for integer factoring.
More recently, classical approaches have been complemented by several quantum protocols. However, they have been subsequently shown to
be inadequate because of susceptibility to different attacks~\cite{Bennett}, or practical difficulties~\cite{BrassardCrep,CrepKil}.

A result of Lo~\cite{Lo} put a damper on the quantum efforts. He showed in 1996 that an ideal solution cannot exist even in the quantum world -- any 
protocol that guarantees perfect concealment of $b$ against Bob actually leaves the database completely vulnerable to 
attacks by Alice.

Since then, several workarounds have been proposed (see e.g.~\cite{Giovannetti}), albeit with some vulnerabilities. 
Recently, Jakobi et al.~\cite{Jakobietal} made interesting progress by proposing a protocol that circumvents the
impossibility proofs at the cost of perfect concealment. Their protocol relies on well known QKD techniques to establish an Oblivious Key between Alice and Bob that fulfills the OT requirements to a large extent while being consistent with Lo's proof. Therefore, we refer to their approach as Quantum Oblivious Key Distribution (QOKD).

The QOKD protocol offers good database security as well
as user privacy and has been shown to be resilient to several attacks. 
However, some problems remain in the communication complexity of their solution. A transfer of
the $N$ qubits is costly in itself, as modern databases are extremely large. The QOKD protocol involves transferring in addition at least $kN$ qubits, where $k$ is a security parameter. It turns out that Alice will on an average get to know about $N(\frac{1}{4})^k$ bits 
of the database. Therefore, unless $k$ also increases with $N$, the number of bits that become known to Alice
in addition to the one she is supposed to know increases with $N$. By the same coin, if it is required to keep this number a constant, 
$k$ will have to rise at least logarithmically with $N$.

\subsection{This Paper \label{subsec:level1}}
In this paper, we first suggest a modification of the initial QOKD protocol wherein the number of qubits that need to be exchanged is reduced to $N$. We then investigate the impact of the modification on database security and user privacy. Subsequently, we show that the modification can be generalized to reduce the required quantum communication complexity even further. We show simple numerical examples of the generalization indicating that at most a few hundred qubits are sufficient even for extremely large databases.

This paper is arranged as follows. The next section gives a brief account of the QOKD protocol of Jakobi et al~\cite{Jakobietal}. In section III we discuss
its modification, analysis and generalization. Section IV concludes the paper.
 

\section{The initial Quantum Oblivious Key Distribution Protocol}
\subsection{Brief Sketch}

The QOKD protocol for SPIR proceeds in three phases: First, a key is established between Bob and Alice using the SARG04~\cite{Scarani2004} Quantum Key 
Distribution (QKD) protocol. 
In the second phase, this key is processed to produce an oblivious key $O\!K$, a string of $N$ bits. While Bob has complete knowledge of this oblivious key,
Alice knows only a few bits conclusively. Note that this $O\!K$ is not perfect and therefore does not contradict Lo's impossibility proof. In the final phase, the oblivious key $O\!K$ is used to classically encrypt the database so that Alice can learn the bit that she is interested in.

\textbf{First phase:} In contrast to BB84, the SARG04 QKD protocol uses the basis to encode a bit. For example, let the ``up-down" basis $\updownarrow$ 
encode bit value 0 
and ``left-right" basis encode 1. 
The protocol would then use the four states $\ket{\uparrow}$, $\ket{\rightarrow}$, $\ket{\downarrow}$, and 
$\ket{\leftarrow}$, with $\left|\langle{\uparrow}\ket{\rightarrow}\right|^2=\frac{1}{2}$ etc.
To establish one bit of the key, Bob prepares one of these states and sends it to Alice. He then announces the sent state and one of the other basis.
For instance, to send 0, Bob can prepare the state $\ket{\uparrow}$ and announce the pair
$\{\ket{\uparrow},\ket{\rightarrow}\}$. Alice then has to determine whether Bob sent $\ket{\uparrow}$ or $\ket{\rightarrow}$. A simple 
way to do this is to measure the received state in one of the two bases and hope for a result that will exclude one of the announced states. 

In the example above, measuring in left-right basis will yield the result $\ket{\leftarrow}$ with probability $1/2$, which excludes the announced state 
$\ket{\rightarrow}$. This allows Alice to conclude that the state sent by Bob must have been $\ket{\uparrow}$. A measurement in the up-down-basis would 
never yield a conclusive result as the only possible result is $\ket{\uparrow}$. 

Since Alice chooses the correct basis half of the time and then obtains a conclusive result
with probability $1/2$, the overall probability of having a conclusive result is $\frac{1}{4}$ in SARG04. Therefore, Alice will know only a quarter of 
the sent bits with certainty; the values of the rest are inconclusive.
Indeed, a ``bit" can now
also have the value ``inconclusive" in addition to 0 or 1.

To proceed with the extraction of the oblivious key $O\!K$, all sent bits are kept 
for the second phase. Note that this procedure is completely loss-independent~\cite{comm-loss}.

\textbf{Second phase:} The steps of the first phase are repeated until a raw key $R$ with elements $\{q_i\}, i=1\ldots kN$ is established. Alice will know the values of 
$\frac{kN}{4}$ bits of $R$ conclusively, while Bob knows all. The problem now is to extract (from the raw key $R$) an oblivious key $O\!K$, a string of 
$N$ bits completely known to Bob but of which Alice only knows a few elements. 
To that end, we form $N$ groups of $k$ qubits each.
The elements of the oblivious key $O\!K$ are then defined as the $\mathrm{X\!O\!R}$ of the $N$ groups: $O\!K_j=\mathrm{X\!O\!R}(q_{k\cdot j},q_{k\cdot j+1},\ldots,q_{k\cdot j+k-1})$ for $1\leq j \leq N$. 
Therefore, even if one of the bits is inconclusive for Alice, her evaluation of $\mathrm{X\!O\!R}$ will be inconclusive. If Alice conducts her measurement as
described in phase I, the probability that Alice knows all the bits of a group conclusively and can therefore compute the parity of the group is $(\frac{1}{4})^k$. Finally, she will 
know on average $N(\frac{1}{4})^k$ elements of the oblivious key $O\!K$ conclusively. $k$ should be chosen such that Alice knows on average only a small number $c$ of key bits, i.e. $k=\log_4(N/c)$. With a probability of $e^{-c}$ Alice is left with no known bit of $O\!K$ and the protocol must be restarted.

\textbf{Third phase:} After completion of the second phase, an oblivious key $O\!K$ is established such that on average $c$ bits are known to Alice, while 
Bob knows $O\!K$ completely. This key is used to bitwise encrypt the database $D\!B$ ensuring that Alice obtains little information besides the bit she is interested in. Supposing Alice knows the bit $O\!K_j$ of the key and is interested in $D\!B_b$, the $b^{th}$ bit of $D\!B$, she communicates the shift $s=j-b$ to Bob. As described in the introduction, Bob then encrypts the database bitwise as $D\!B'_a = \mathrm{X\!O\!R}(D\!B_a,O\!K_{a+s})$, $1\leq a\leq N$, announces the encrypted database $D\!B'$, and Alice recovers the bit that she wanted: $D\!B_b=\mathrm{X\!O\!R}(D\!B'_b,O\!K_j)$. 
If $\{j'\}$ are the indices of the $c-1$ other bits in $O\!K$ that she learns conclusively after phases I and II, she can also get to know some more bits $D\!B_{b'}=\mathrm{X\!O\!R}(D\!B'_{b'},O\!K_{j'})$ of the database. However, the $\{j'\}$ are randomly distributed in the $O\!K$ and will generally not allow Alice retrieving a second bit of interest to her.

\subsection{On the Security of Quantum Oblivious Key Distribution}

Jakobi et al~\cite{Jakobietal} provide interesting arguments for the security of their QOKD scheme, while studying the most obvious attacks directly. Like all quantum SPIR protocols, QOKD cannot offer perfect security for both sides but exploits a trade-off between database security and user privacy.

\textbf{Database security:} At the outset, Jakobi et al~\cite{Jakobietal} let us know that the above protocol actually provides Alice with information on inconclusive bits, too. In the example of phase I, Alice measuring Bob's sent state $\ket{\uparrow}$ in the up-down basis will never yield a conclusive result as it is not possible to rule out any of the two announced states $\{\ket{\uparrow},\ket{\rightarrow}\}$. However, with this measurement, Alice will always find the state $\ket{\uparrow}$. Having chosen the same measurement basis as Bob used for state preparation, Alice will always find his sent state ``inconclusively". As this happens half of the time, and as the other inconclusive result ($\ket{\rightarrow}$ in the example) is found with only $\frac{1}{4}$, Alice has indeed a guess on which state Bob had sent. By assuming her ``inconclusive" outcome is actually the state he had prepared, she will be correct about the bit value with likelihood $\frac{2}{3}$. This partial information will be washed out during the extraction of the $O\!K$ in such a way that in a group where Alice measures all but $x$ bits conclusively, she will guess the key bit correctly with $\frac{3^x+1}{2\cdot3^x}, x\geq 1$.

While analyzing the protocol's security, one must assume in general that Alice has a quantum memory at her disposal and is able to postpone her measurements until after Bob's SARG04 state pair announcement. She then knows that her measurement must distinguish, for instance, the states $\ket{\uparrow}$ and $\ket{\rightarrow}$ in order to decipher the sent bit. It can then be shown that Alice can perform an unambiguous state discrimination (USD) measurement which is successful with a probability of at most $p_{USD}=1-F(\ket{\uparrow},\ket{\rightarrow})=1-1/\sqrt{2}\approx 0.29$, where $F$ is the fidelity. If Alice measures each received qubit individually, this attack is optimal and Alice will have on average $0.29N$ conclusive qubits instead of $0.25N$ before starting phase II of the protocol. However, this fact has only limited impact as it will increase the number of key elements known to Alice by only $\left(\frac{p_{USD}}{0.25}\right)^k\approx(1.16)^k$, where typically $k<10$.

Instead of performing individual measurements, Alice can also perform a joint measurement on $k$ qubits in order to directly measure their overall parity. This way, she directly measures the associated key bit without using individual bit values.
Jakobi et al~\cite{Jakobietal} show that the success probabilities for USD as well as Helstrom maximal information gain measurements on $k$-qubit states decline rapidly with increasing $k$.
Therefore, Alice's knowledge on the final key is physically restricted by the impossibility to perfectly discriminate the non-orthogonal states used for encryption of the key elements.

\textbf{User privacy:} Jakobi et al.~\cite{Jakobietal} argue that Bob is able to obtain limited information on the conclusiveness of Alice's bits but will then lose information on which bit value she has actually measured. He will thus introduce errors. For example, sending $\ket{\nearrow}$ or $\ket{\swarrow}$ while announcing a pair
$\{\uparrow,\rightarrow \}$ will yield a probability for Alice to measure conclusively of $p_{-}=\frac{1}{2}-\frac{1}{2\sqrt{2}}\approx 0.15$ or $p_{+}=\frac{1}{2}+\frac{1}{2\sqrt{2}}\approx 0.85$, respectively. This turns out to be optimal -- Bob can bias the conclusiveness probability $p$ for Alice's qubits within the limits $p_{-}\leq p \leq p_{+}$. At the same time, sending $\ket{\nearrow}$ or $\ket{\swarrow}$ will obviously not give Bob any information on the result of Alice's measurement. 
In fact, Bob cannot know the measurement basis Alice chose, which implies that it is impossible for him to have both increased information on her conclusiveness and full information on the bit value she measures (if conclusive). Every manipulation will hence create errors in the oblivious key. 

These characteristics are a consequence of the use of non-orthogonal states in SARG04 and the no-signaling principle.
As a consequence, the protocol exploits fundamental physical principles to ensure database security and user privacy while allowing small additional information gains for both sides thus preventing a conflict with Lo's impossibility proof.

\subsection{The Problem of Efficiency}
The number $c$ of bits revealed to Alice at the end of SARG04 and $\mathrm{X\!O\!R}$ing of the $N$ groups of $k$ bits is on average
$N(\frac{1}{4})^k$. Thus, unless $k$ increases with $N$, 
$c$ would also increases with $N$. In particular, $k$ needs to increase at least logarithmically
with $N$ to ensure that $c$ remains constant and quantum communication complexity is therefore $O(N \log N)$.
Given the size of modern databases (which run into petabytes), such an increase should be avoided as this would be far too costly for the communication of only one bit to Alice.

We now show that it is possible to reduce the required quantum communication complexity, first to $O(N)$ and subsequently even below, while maintaining the protocol's security.

\section{The Road to Communication-efficiency}

\subsection{The Modified Protocol \label{sec:sketch}}

We propose modifying the second phase of the above protocol in such a way that every element of $R$ is replaced by the $\mathrm{X\!O\!R}$ of its current value with the 
values of the $k-1$ elements immediately following it. The last element is replaced by the $\mathrm{X\!O\!R}$ of its current value with the value of the first elements.

Then, the modified protocol is as follows:
\begin{itemize}
\item Let $R$ be the raw key after execution of SARG04 for $N$ bits. Then, while Bob knows the entire $R$, on average three quarters of the elements of $R$ are inconclusive at Alice's end.
\item Define the elements of $O\!K$ as follows: $O\!K_j=\mathrm{X\!O\!R}(q_j,\ldots,q_{j+k-1})$ for $j=1\ldots N$ (with $q_{N+x}:=q_x$ for $1<x<k-1$).
\item If no bit survives at Alice's end, repeat the above two steps.
\item Continue with the steps of the third phase for database exchange and verification.
\end{itemize}

The modification that we have just described requires a quantum communication complexity of $N$ and is based on the following observations.

Suppose we have a coin that shows head with probability $p$ and tails with probability
$1-p$ when tossed. It is a folklore theorem that when such a coin is tossed $N$ times,
the length of the longest streak of heads is $\Theta(\log_{1/p} N)$ with high probability~\cite{Cormen2001}. 

The analogue of a streak of heads in coin tosses is a streak of conclusively known
bits at the end of the SARG04 protocol for $N$ qubits. Tails would therefore be
analogous to the inconclusive bits.

We will now argue that with high probability, the expected number of times such a maximum length streak occurs is $O(1)$.
Let a bit be conclusively revealed to Alice with probability $p$. Then, the probability
of a contiguous streak of conclusively revealed $l$ bits is $p^l$.
Let $X_{il}$ be the indicator random variable that takes the value 1 if a streak of length $l$ starts at position
$i$ in the key and 0 otherwise.
Thus, $X_l=\sum_{i=1}^{N}X_{il}$ is the random variable that counts the number of streaks of
length $l$ in the key. By linearity of expectation, the expected number of streaks of length $l$ is
$\sum_{i=1}^{N}E[X_{il}]$. Given that $Pr[X_{il}=1]=p^l$, we have $E[X_l]= \sum_{i=1}^{N}p^l$.
That is, $E[X_l]=Np^l$. For $l=\log_{1/p}N$, we have $E[X_l]=1$. 
Moreover, by Markov inequality, the probability that the number of such streaks 
exceeds some $t$ is at most $\frac{E[X_l]}{t}$.

For instance, if we take $p=\frac{1}{4}$ and $l=k=\log_{4}(N/c)$, we find $E[X_k]=c$. That is, the above 
procedure will yield on average $c$ streaks of length $k$, where $k=\log_4(N/c)$ as in the original QOKD protocol.
Finally, by Markov inequality, the probability that the number of such streaks exceeds $E[X_k]^m=c^m$, for any $m> 1$, is at most $\frac{1}{c^{m-1}}$.

In other words, it is likely that (i) there is at least one streak of length $k$ in the key,
(ii) there is only a small number $c$ of streaks of length $k$, and 
(iii) every other streak in $O\!K$ is less than $k$ in length. 

We report in table~\ref{tab:simulations} simulations that justify the protocol.
As pointed out by Jakobi et al.~\cite{Jakobietal}, even with a quantum memory, Alice can conclusively obtain only about $0.29$ of the bits after execution of SARG04 (the first step of the protocol). For this reason, and in continuation of our running example, we run the simulations on $p=\frac{1}{4}$ and $p=1-\frac{1}{\sqrt{2}}$ respectively with the same $k$ for both.

\begin{table}[tc]
\begin{center}
\begin{tabular}{c | c c c c c c }
$N$	    	& $10^4$ & $10^5$ & $10^6$ & $10^7$ & $10^8$ \\	
\hline
k	    	& $6$ & $7$ & $9$ & $11$ & $13$  \\

At least one	&  $81$ & $98$ & $95$ & $86$ & 79\\
Average $p=0.25$& $2.37$ & $6.5$ & $4.09$ & $2.45$ & 2.17 $$\\
Average $p=0.29$& $6.46$ & $18.9$ & $15.45$ & $13.42$ & 15.74$$\\
\end{tabular}
\caption{Simulation over 100 runs of the modified QOKD protocol with database size $N$. ``Average" denotes the average number of survivors and ``At least one" denotes the number of runs that have at least one survivor.}
\label{tab:simulations}
\end{center}
\end{table}

\subsection{On the Security of the modified protocol}

The security considerations of~\cite{Jakobietal} presented above largely apply to the modified protocol as well as the changes only concern the post-processing and 
extraction of the oblivious key.

\textbf{Database security:} If Alice has a quantum memory at her disposal, she is able to postpone her measurement after the state announcement of Bob during the SARG04 phase. As discussed before, when measuring each received qubit individually, this attack is optimal and directly covered by the considerations on the likelihood of conclusive streaks in section \ref{sec:sketch} using $p_{USD}=0.29$ instead of $p=0.25$. The impact is precisely as before an increase in known key bits for Alice by a factor of $\left(\frac{p_{USD}}{0.25}\right)^k\approx(1.16)^k$.
However, while individual key bits are as hard to extract as in the initial protocol, the modified version offers less protection with respect to the relative parities between key bits. The difference between two consecutive key bits $O\!K_j$ and $O\!K_{j+1}$ consists in the substitution of the qubit $q_j$ by $q_{j+k}$, i.e. $O\!K_j=\mathrm{X\!O\!R}(q_j,q_{j+1},...,q_{j+k-1})$ and $O\!K_{j+1}=\mathrm{X\!O\!R}(q_{j+1},q_{j+2},...,q_{j+k-1},q_{j+k})$, where the $q_i$ are the bit values corresponding to Bob's sent states. The parity of $O\!K_j$ and $O\!K_{j+1}$ is revealed upon successful measurement of $q_j$ and $q_{j+k}$. We note hence that the reduction in communication complexity comes at the cost that parity information is easier to obtain.

With respect to joint measurement, the new aspect in the modified protocol is that each qubit contributes to $k$ different key elements. Looking at a key bit $O\!K_j=\mathrm{X\!O\!R}(q_j,q_{j+1},..,q_{j+k-1})$, we can assume without loss of generality that Bob announces for all $k$ qubits a SARG04 pair of $\{\uparrow,\rightarrow \}$. The initial state before Alice's measurement of $O\!K_j$ is then $\rho_k =\frac{1}{2^k} \bigotimes_{i=j}^{j+k-1} (\ketbra{\uparrow}_i+\ketbra{\rightarrow}_i)$. Alice now performs a joint USD measurement on $\rho_k$ in order to retrieve $O\!K_j$ directly.
This USD measurement can either be conclusive or non-conclusive, with conclusive results being increasingly unlikely with higher $k$~\cite{Jakobietal}. In case of a conclusive outcome, Alice will know the overall parity $O\!K_j$ of $\rho_k$, and the state after the measurement is given by all possibilities with parity $O\!K_j$: $\rho_k^{O\!K_j} =\frac{1}{2^{k-1}} \sum_{\mathrm{X\!O\!R}(q_j,..,q_{j+k-1})=O\!K_j} \ketbra{q_j,...,q_{j+k-1}}$. Assuming Bob announced a SARG04 state pair $\{\uparrow,\rightarrow \}$, $\ket{q_i}$ should be read as $\ket{q_i=0}=\ket{\uparrow}$ and $\ket{q_i=1}=\ket{\rightarrow}$; that is, the states are not orthogonal.
Alice can now try to determine the parity of the next key element $O\!K_{j+1}=\mathrm{X\!O\!R}(q_{j+1},..,q_{j+k-1},q_{j+k})$. Since all but one of these qubits are part of $\rho_k^{O\!K_j}$, realizing the measurement of $O\!K_{j+1}$ implies tracing out the qubit $q_{j}$ from $\rho_k^{O\!K_j}$, which simply yields $Tr_j \rho_k^{O\!K_j}=\frac{1}{2^{k-1}}\bigotimes_{i=j+1}^{j+k-1} (\ketbra{\uparrow}_i+\ketbra{\rightarrow}_i)$, the initial SARG04 state before measurement for a $k-1$ qubit state.
All parity information is hence erased from this sub-state and measuring $O\!K_{j+1}$ is exactly the same (difficult) task as measuring $O\!K_{j}$.

Now we consider the case of Alice's joint USD measurement being inconclusive. Per definition, the parity of the $k$ qubits' ensemble is lost and can no longer be retrieved.
That is, depending on the concrete design of the measurement, at least one of the $k$ qubits must have lost its bit value information and can no longer be used to define other key elements.
As each qubit contributes to $k$ different key elements, Alice's failed joint USD measurement of a single key element renders in fact the decoding of $k$ key elements impossible.
In this sense, our modification can indeed increase database security.

\textbf{User privacy:} The fundamental arguments of~\cite{Jakobietal} for user privacy were based on the
impossibility of perfectly distinguishing non-orthogonal quantum states and superluminal communication.
These remain valid for the modified protocol as well. In particular, we remind the reader that Bob has no measurement that would allow him learning both conclusiveness and Alice's bit value information.
Our first observation is that by manipulating the conclusiveness of a single qubit $q_i$, Bob will impact the conclusiveness probability of the $k$ key elements that use $q_i$. 
However, the same is true for the error he introduces which also affects $k$ key elements and becomes hence easier to detect.
A possible strategy for Bob to narrow down Alice's conclusive bits is to increase the conclusiveness of a (contiguous) part of his sent qubits while reducing it for the rest. Remembering that $p_{+}p_{-}=\frac{1}{2}$, increasing the conclusiveness of $p_{-}N$ qubits to $p_{+}$ while reducing the conclusiveness of the remaining $p_{+}N$ qubits to $p_{-}$ will maintain Alice's statistics of conclusive bits in $R$.
Neglecting border effects, these two parts can be seen as independent strings on which the results of section \ref{sec:sketch} can be applied. For the number of streaks of length $k$ one finds $E_{+}=p_{-}Np_{+}^k$ and $E_{-}=p_{+}Np_{-}^k$. It follows that $\frac{E_{+}}{E_{-}}=\left(\frac{p_+}{p_-}\right)^{k-1}\gg 1$. Therefore, Bob knows that the conclusive bit, which Alice will use to code the database element she is interested in, will lie with a high probability of $\frac{E_+}{E_+ + E_-}$ in the high conclusiveness part of $O\!K$. However, we note the following observations: (1) Bob's knowledge remains considerably limited as $p_{-}N\approx 0.15 N$ key elements are still equally likely, (2) Bob does not know a single bit of the final key correctly and will thus give completely random answers during the third phase, and (3) Alice will have significantly more strings of length $k$ than expected since $E_{+} \gg N\left(\frac{1}{4}\right)^{k}$, which should make her more than suspicious. Indeed, as the protocol is both linear in $p$ (number of conclusively measured qubits) as well as non-linear (number of streaks of length $k$), Bob altering the conclusiveness of qubits systematically will easily show in Alice's statistics.

\subsection{Generalization}

In the present modification of the QOKD protocol, a bit of the final key is defined as the parity of a streak of $k$ qubits $O\!K_j=\mathrm{X\!O\!R}(q_j,q_{j+1},...,q_{j+k-1})$. The reduction in communication complexity arises from the re-utilization of each qubit as a contributing element for $k$ bits of the final key, i.e. qubit $q_j$ is used in the definition of the key bits $O\!K_{j-k+1}$ to $O\!K_{j}$. This idea can be generalized in order to further reduce communication requirements: Let us assume phase I of the QOKD protocol is performed until $M<N$ qubits are distributed to Alice. In order to define the elements of the oblivious key, we now consider all possible combinations of $k$ out of these $M$ qubits. This allows to extract a key of length $\binom{M}{k}$ as each combination constitutes an independent parity functions of $k$ qubits in the sense introduced by our modified protocol. As such, by considering all these possible definitions of key bits, the minimal quantum communication complexity required for a $N$-bit database is given by $\binom{M_{min}}{k} \geq N$.

Table \ref{tab:generalization} provides some numerical examples of the impact of the discussed generalization. As can be seen, in this generalization, there is a certain freedom in choosing $M$ and $k$. While high $k$ and small $M$ will increase database security but also increase the abortion probability, low $k$ and high $M$ achieve the opposite.
Even for huge databases, the required quantum communication complexity can be reduced to under 100.

However, the small number of exchanged qubits gives rise to generally poor statistics making  statistical analyses somewhat unreliable. Also, as this generalization presents an extreme case of re-using qubits for key definition and hence for reduction in quantum communication complexity, it does not come as a surprise that security is considerably less tight. Whereas in the initial and in the modified protocol a small constant number of database bits $c=Np^k$ was revealed to Alice on average, the generalized protocol provides Alice with significantly more bits, especially if the abortion probability should be low ($pM\geq k$). For instance, if Alice measures $k+x$ of the $M$ qubits conclusively, she is able to calculate $\binom{k+x}{k}$ key elements. Additionally, even when Alice measures only exactly $k$ qubits conclusively and can hence only calculate one single key bit, she is still able to calculate parities between many key elements.
As such, the generalized protocol provides only little database security. Fortunately, the protocol is sufficiently cheap to be re-performed a couple of times, which allows to completely re-establish database security as we will see in the next section.

\begin{table}[tc]
\begin{center}
\begin{tabular}{c | c c c c c }
&&&$N\geq10^5$\\
\hline
$M_{min}$	    	&$41$		&$29$		&$23$		&$21$	 	 &$20$\\
$k$ 	    	&$4$		&$5$		&$6$		&$7$ 		 &$8$\\
 
Average  	&$397$		&$131$		&$46$		&$28$ 	 	 &$19$\\

No bit	&$3.8\%$	&$11.5\%$	&$46.8\%$	&$74.4\%$	 &$89.8\%$\\
\end{tabular}\\
\begin{tabular}{c | c  c  c  c  c }
&&&$N\geq10^{10}$\\
\hline
$M_{min}$	    	&$71$		     &$58$		&$50$		&$45$		&$42$\\

$k$ 	    	&$8$ 		     &$9$		&$10$		&$11$		&$12$\\
 
Average 	&$162531$	     &$41833$		&$11714$	&$4094$		&$1876$\\

No bit	&$1.2\%$	     &$2.9\%$		&$16.4\%$	&$40.9\%$	&$64.9\%$\\

\end{tabular}\\
\caption{Calculated examples for generalized QOKD of databases of size $10^5$ and $10^{10}$ for different combinations of quantum communication complexity $M$ and security parameter $k$. ``Average" denotes the average number of survivors conditioned on cases with at least one survivor and ``No bit" the probability for no survivors. \label{tab:generalization}}
\end{center}
\end{table}

\subsection{Enhancing database security}

It is possible to significantly enhance database security by re-performing $r$ times either of the presented variants of QOKD as follows: in each of the $r$ rounds an oblivious key is generated, $O\!K^i, i=1\ldots r$. 
To obtain the final key $O\!K^\text{fin}$, Alice is asked to combine these $r$ keys bitwise with relative shifts $s_i$ she can freely choose: $O\!K^\text{fin}_j=\displaystyle\bigoplus\limits_{m=1}^r O\!K^m_{j+s_m}$.
This final key is then used to encrypt the database as described in phase III of the original protocol.
This procedure serves the following purpose. Using QOKD to generate the $r$ oblivious keys ensures that Alice only has partial knowledge on each of them. 
Therefore, combining them will further reduce her knowledge on the key while the free choice of the offset ensures that Alice always retains at least one element of the sum string.
For instance, let us look at the first case of table \ref{tab:generalization} with $r=2$: Alice generates two keys of $10^5$ bits, of which she knows $400$ conclusively each. It is important to remember that these bits are randomly distributed over the key strings.
As such, just combining these strings without choosing the optimal offset will yield on average $1.6$ remaining conclusive bits.
Numerical simulations show that by selecting the optimal offset, Alice will be able to retain on average $9.7$ known bits of the sum string. Obviously, $r=3$ will further reduce Alice knowledge. In principle, choosing a large $r$ will almost guarantee that Alice retains one and only one bit in the end \cite{Lo-argument}. Note that this procedure will also erase parity information that Alice can gather in the protocols proposed in this paper.

The presented ``dilution process'' can obviously obviously ensure adequate database security and allows hence to take full advantage of the achieved reduction in quantum communication complexity.

\section{Conclusion}

We showed that the protocol proposed by Jakobi et al. can be modified to reduce the required quantum communication complexity without compromising its security and while maintaining its strength of loss-resistance, practical feasibility, and integrability with current QKD devices. As a consequence, it is now possible to bring very large databases into the scope of Quantum Oblivious Key Distribution. Moreover, the modified protocol is sufficiently cheap in terms of quantum communication complexity so as to construct approximate versions of a whole range of quantum cryptographic algorithms based on SPIR. As such, Quantum Oblivious Key Distribution can significantly add to what can practically be realized today in the realm of Quantum cryptography and, together with QKD, it might well provide the basis for all practical future applications of quantum cryptography.


\begin{thebibliography}{100}

\bibitem{Kilian} Joe Kilian.
\newblock Founding cryptography on oblivious transfer.
\newblock In {\em Proceedings of the twentieth annual ACM symposium on Theory of computing}, STOC '88, pages 20--31, New York, NY, USA, 1988. ACM.

\bibitem{Rabin} Michael~O. Rabin.
\newblock How to exchange secrets by oblivious transfer.
\newblock {\em Technical Report TR-81, Aiken Computation Lab, Harvard University}, 1981.

\bibitem{EGL} Shimon Even, Oded Goldreich, and Abraham Lempel.
\newblock A randomized protocol for signing contracts.
\newblock {\em Commun. ACM}, 28:637--647, June 1985.

\bibitem{Crepeau} Claude Cr\'{e}peau.
\newblock Equivalence between two flavours of oblivious transfers.
\newblock In {\em A Conference on the Theory and Applications of Cryptographic Techniques on Advances in Cryptology}, CRYPTO '87, pages 350--354, London, UK, 1988. Springer-Verlag.

\bibitem{OK} D. Beaver, {\it Lecture Notes in Computer
Science, Vol. 963} (Springer, London, 1995), p. 97.

\bibitem{jacmChor} Benny Chor, Eyal Kushilevitz, Oded Goldreich, and Madhu Sudan.
\newblock Private information retrieval.
\newblock {\em J. ACM}, 45(6):965--981, 1998.

\bibitem{Bennett}
Charles~H. Bennett, Gilles Brassard, Claude Cr\'{e}peau, and Marie-H\'{e}l\`{e}ne Skubiszewska.
\newblock Practical quantum oblivious transfer.
\newblock In {\em Proceedings of the 11th Annual International Cryptology Conference on Advances in Cryptology}, CRYPTO '91, pages 351--366, London, UK, 1992. Springer-Verlag.

\bibitem{BrassardCrep} Gilles Brassard and Claude Cr\'{e}peau.
\newblock Quantum bit commitment and coin tossing protocols.
\newblock In {\em Proceedings of the 10th Annual International Cryptology Conference on Advances in Cryptology}, CRYPTO '90, pages 49--61, London, UK, 1991. Springer-Verlag.

\bibitem{CrepKil} C.~Cr\'{e}peau and J.~Kilian.
\newblock Achieving oblivious transfer using weakened security assumptions.
\newblock In {\em Proceedings of the 29th Annual Symposium on Foundations of Computer Science}, pages 42--52, Washington, DC, USA, 1988. IEEE Computer Society.

\bibitem{Lo} Hoi-Kwong Lo.
\newblock Insecurity of quantum secure computations.
\newblock {\em Phys. Rev. A}, 56(2):1154--1162, Aug 1997.

\bibitem{Giovannetti} Vittorio Giovannetti, Seth Lloyd, and Lorenzo Maccone.
\newblock Quantum private queries: security analysis.
\newblock {\em IEEE Trans. Inf. Theor.}, 56:3465--3477, July 2010.

\bibitem{Jakobietal} Markus Jakobi, Christoph Simon, Nicolas Gisin, Jean-Daniel Bancal, Cyril Branciard, Nino Walenta, and Hugo Zbinden.
\newblock Practical private database queries based on a quantum-key-distribution protocol.
\newblock {\em Phys. Rev. A}, 83(2):022301, Feb 2011.

\bibitem{Scarani2004} V.~Scarani, A.~Acin, G.~Ribordy, and N.~Gisin.
\newblock Quantum cryptography protocols robust against photon number splitting attacks for weak laser pulse implementations.
\newblock {\em Physical Review Letters}, 92(5):057901, 2004.

\bibitem{comm-loss} Transmission and detection losses can easily compromise a protocol's security, see for instance \cite{Giovannetti}. The receiver, in our case Alice, could perform her measurement and, if the measurement outcome is inconclusive, claim that the qubit was lost. In particular, if the receiver only pretends the presence of losses, she can to a large extend pick convenient measurement results and thus increase her knowledge. Fortunately, the timing of the protocol can easily prevent this: First, Bob sends a qubit to Alice. She then performs her measurement and acknowledges it if her detection was successful, otherwise a new qubit is sent. Only following this confirmation, Bob will announce the SARG04 state pair. As without this information, she is unable to evaluate if her measurement was conclusive or not, the impact of losses on the protocol's security is eliminated. Please note that a quantum memory will not change this situation.

\bibitem{Cormen2001} T.~H. Cormen, C.~E. Leiserson, R.~L. Rivest, and C.~Stein.
\newblock {\em Introduction to Algorithms}.
\newblock The MIT Press, New York, 2001.

\bibitem{Lo-argument} Please note that no direct conflict with Lo's impossibility proof arises. In his proof, Lo assumes (1) perfect concealment of Alice's choice $b$ against Bob and argues that Alice can then read the entire database (2) if the states representing the different database elements are eigenstates of her measurement operator. Both assumptions are not fulfilled in this limit.

\end{thebibliography}
\end{document}